\documentclass[showpacs,floatfix,twocolumn]{revtex4}
\usepackage{graphicx}

\makeatletter
\newbox\slashbox \setbox\slashbox=\hbox{\large$/$}
\def\pslash#1{\setbox\@tempboxa=\hbox{$#1$}
  \@tempdima=0.5\wd\slashbox \advance\@tempdima 0.5\wd\@tempboxa
  \copy\slashbox \kern-\@tempdima \bo v  x\@tempboxa}

\makeatother
\newcommand{\be}{\begin{eqnarray}}
\newcommand{\ee}{\end{eqnarray}}
\def\beq{\begin{equation}}
\def\eeq{\end{equation}}
\def\SB{S$\chi$SB}

\begin{document}

\title{Chiral phase transition in lattice QCD as a metal-insulator transition}
\author{Antonio M. Garc\'{\i}a-Garc\'{\i}a}
\affiliation{Physics Department, Princeton University, Princeton,
New Jersey 08544, USA}
\affiliation{The Abdus Salam International Centre for Theoretical
Physics, P.O.B. 586, 34100 Trieste, Italy}
\author{James C. Osborn}
\affiliation{Physics Department \& Center for Computational Science,
 Boston University, Boston, MA 02215, USA}

\begin{abstract}
We investigate the lattice QCD Dirac operator with staggered fermions 
 at temperatures around the chiral phase transition. 
We present evidence of a metal-insulator transition in the low lying modes of
 the Dirac operator around the same temperature as the chiral phase transition.
This strongly suggests the phenomenon of Anderson localization 
 drives the QCD vacuum to the chirally symmetric phase in 
 a way similar to a metal-insulator transition in a disordered conductor.
We also discuss how Anderson localization affects the usual phenomenological 
 treatment of phase transitions {\it a la } Ginzburg-Landau.
\end{abstract}
\maketitle

One of the most important features of the infrared limit of the strong
 interactions is the spontaneous breaking of the approximate chiral
 symmetry.
The order parameter associated with this spontaneous chiral symmetry breaking
 (\SB{}) is the chiral condensate, $\langle {\bar \psi} \psi \rangle$,
 which in the absence of \SB{} would vanish as the quark mass goes to zero.
In nature the lightest quarks are not massless so a nonzero 
 condensate is expected even in a free theory.
However the small quark mass can only account for a small percentage
 of the chiral condensate, the rest has its origin in the strong
 non-perturbative color interactions of QCD.
Although lattice simulations have already provided overwhelming evidence
 that \SB{} is a feature of QCD it is still highly desirable
 to understand its origin in more simple terms. 

Simplified models of QCD where gauge configurations are given 
 by instantons have played a leading role in the description of the 
 \SB{} \cite{callan,diakonov,shuryak}.
Instantons \cite{polyakov}, 
 originally introduced in QCD by 't Hooft \cite{thooft} to 
 solve the so called $U(1)$ problem, are classical 
 solutions of the Euclidean Yang-Mills equations of motion.
Their relation with the \SB{} 
 stems from the fact that the QCD Dirac operator has an
 exact zero eigenvalue in the field of an instanton. 
In the QCD vacuum, these zero modes coming from different 
 instantons mix together to 
 form a band around zero.  
It turns out the spectral density, $\rho(\lambda)$,
 of the Dirac operator in this region is directly related to 
 the  chiral condensate   
 through the Banks-Casher
 relation \cite{bank},
\be
\langle {\bar \psi} \psi \rangle &=&
 - \lim_{m \to 0} \lim_{V \to \infty} \int_{0}^{\infty}
 \frac{2m}{m^2+\lambda^2}  \frac{\rho(\lambda)}{V} d\lambda \nonumber \\
&=& - \lim_{\lambda \to 0} \lim_{V \to \infty} \frac{\pi \rho(\lambda)}{V} ~,
\label{BC}
\ee
 where $V$ is the space-time volume.
Based on this result it was shown that the instanton contribution was capable
 of producing a nonzero chiral condensate with a value close to the
 phenomenological one \cite{shuryak,diakonov}
 (see \cite{SS97} for a detailed review). 

For sufficiently high energies the non-abelian gauge interaction of QCD 
 is weak (asymptotic freedom) and chiral symmetry is restored.
This poses an interesting question:
 How is the chiral symmetry restored as we go from low to high energies or,
 equivalently, from low to high temperatures?

A standard approach to the chiral phase transition
 is to invoke universality arguments \cite{wilczek}, namely, it is assumed
 that the nature of the transition is determined solely by
 symmetries of QCD.
The chiral phase transition is then studied
 by looking at the most general renormalizable Ginzburg-Landau Lagrangian
 with the chiral symmetries of QCD.
By using a perturbative renormalization group analysis,
Pisarsky and Wilczek \cite{wilczek} found that for two massless flavors
 the transition is expected to be second order.
However, for three or more flavors, it was conjectured to be first order
 due to the absence of infrared stable fixed points in the 
 renormalization group equations. 
The order and the very existence of the chiral transition is also sensitive to
 details such as the mass of the light quarks and whether the above mentioned
 $U(1)$ symmetry is restored at the same temperature as the chiral one.

Generally speaking it is still under debate to what extent effective models
 only based on universality arguments provide an accurate description of
 the chiral phase transition \cite{revchi} (see e.g. \cite{mendes} for 
 a recent discussion of some lattice results).

For instance, it is unclear whether the $\epsilon$ expansion used to determine
 the fixed points is reliable at $\epsilon = 1$ and additionally whether
 nonperturbative effects may alter this behavior completely.

In this paper we present evidence that the phenomenon of Anderson localization
 \cite{anderson} plays a
 crucial role in the chiral restoration.
We suggest that localization drives the system to the chirally symmetric phase
 in a way similar to a metal-insulator transition (also referred to as Anderson
 transition (AT)) in a disordered conductor.
Anderson localization thus counterbalances the effect of the QCD interactions 
 which tend to keep the \SB{} phase.
For the sake of completeness we briefly review some basic facts about
 disordered systems with special emphasis on the phenomenon of Anderson
 localization (see section VI of \cite{guhr} for a review).

Localization properties of a disordered system, i.e. a free particle in a
 random potential \cite{anderson}, are investigated by looking at the
 eigenstates or, more economically, by studying level statistics of
 the Hamiltonian.
In two and lower dimensions, destructive interference caused by backscattering
 produces exponential localization of the eigenstates in real space for any
 amount of disorder.
As a consequence, quantum transport is suppressed, the spectrum is
 uncorrelated (Poisson statistics) and the system becomes an insulator.
In more than two dimensions there exists a metal-insulator transition for
 a critical amount of disorder due to the interplay of destructive 
 interference and tunneling.
By critical disorder we mean a disorder such that, if increased, all the
eigenstates become exponentially localized. For a disorder
strength below the critical one, the system has a mobility edge at
a certain energy which separates localized from delocalized
states. Its position moves away from the band center as the
disorder is decreased. Delocalized eigenstates, typical of a
metal, are extended through the sample and their level statistics
agree with the random matrix theory (RMT) prediction for the appropriate
symmetry.
We note that in the case of QCD the role of Hamiltonian is played by the
 Dirac operator.

Localization has already been investigated in lattice QCD 
 \cite{locgw,locstag,locother}.
The low lying modes of lattice QCD with overlap fermions
 at zero temperature are generally found to be localized even though the
 eigenvalue spectrum agrees with the random matrix prediction \cite{locgw}.
This gives an apparent contradiction since the RMT also predicts extended
 eigenstates.
Conversely, simulations with staggered fermions do find extended states along
 with an eigenvalue spectrum that agrees with RMT \cite{locstag},
 though the localization properties of the low modes in unquenched simulations
 have not been studied in detail near the chiral phase transition.
In the context of instanton liquid models (ILM) at zero 
 temperature, the \SB{} has been related
 to the conductivity (delocalization) in a disordered medium 
 \cite{diakonov1,VO,zahed,aj}.
At nonzero temperature, we have recently reported \cite{aj1} that,
 in agremeent with a previous suggestion \cite{diakolet},
 the chiral phase transition in the ILM is induced by an
 AT of the lowest lying eigenmodes of the Dirac operator.
In this letter we show that the close relations between
 Anderson localization and the chiral phase transition found in the ILM holds
 in full lattice QCD as well.

In the next section we introduce the lattice simulations used in this paper
 and present evidence of a localization transition in quenched lattice QCD.
Then we show that this metal-insulator transition  
 occurs around the same temperature as the chiral restoration.
In section \ref{secunq} we repeat the analysis in full (unquenched)
 lattice QCD with similar results.
The last section provides some further clarifications about the 
relation between Anderson localization and chiral restoration including
 the effect of localization on the standard picture of
 phase transitions {\it a la} Ginzburg-Landau.

\section{Localization of the QCD Dirac operator at nonzero temperature}

In this section we investigate how the localization
 properties of the QCD Dirac operator depend on temperature.
Specifically we are looking for a critical temperature at which
 a metal-insulator transition occurs.
We shall also investigate if the spectral and eigenfunction properties
 around the critical region are compatible with those
 of a disordered conductor at the metal-insulator transition.  

Our first task is to obtain both the eigenvalues and eigenvectors of
 the Dirac operator for different temperatures and then find the
 location, if any, of the metal-insulator transition.
In principle one should also determine the spectral region in which
 the transition takes place.
In the quenched case, according to the Banks-Casher relation, the
 physics of the \SB{} and its eventual restoration is exclusively
 linked with the lowest eigenmodes of the Dirac operator so we will
 look for the metal-insulator transition only in this region.
For dynamical fermions the situation is different, the condensate
 depends on a wider spectral region, so we will examine how much the
 region in which the AT occurs overlaps with the one relevant for the
 chiral condensate.

\subsection{Details of the lattice QCD simulations}

For the present study we have used quenched and 2+1 flavor lattices
 at couplings ($\beta$) around the chiral restoration transition.
Both sets of lattices were generated using
 a one-loop Symanzik improved gauge action \cite{sym}.
The lattice sizes used are $L^3 \times 4$ where $L$ is the spatial size and
 the number of time slices is always $4$.
The quenched lattices used $L=16$ and $20$ while the unquenched lattices
 used $L=12$ and $16$.
The 2+1 flavor lattices were generated using the $a^2$ tadpole improved
 ``asqtad'' staggered Dirac operator \cite{asqtad} with 2 light quark
 flavors and 1 strange quark flavor with the light quark masses one tenth the
 strange quark mass.
The $L=12$ lattices come from the MILC collaboration and are a subset
 of the ones mentioned in \cite{milclat}.
The $L=16$ lattices were generated using the same parameters as
 the smaller lattices except for the volume.

We have obtained
the lowest 64 eigenvalues and eigenvectors of the asqtad staggered Dirac
operator on these lattices.
In the quenched lattices we first rotated the gauge fields
by an appropriate $Z_3$ transformation so that the Polyakov loop is in
the ``real'' phase.
This avoids having to deal with seeing separate transitions for the different
 phases at different temperatures, although it would be interesting to
 examine the other phases too.
We note that the staggered Dirac operator actually represents four
 copies of the continuum Dirac operator (called ``tastes'') that are mixed
 together. However, since the lattice spacing is large, the scale for the taste breaking
 is much larger than the eigenvalues being studied here.
It is thus expected that the eigenvalues behave as if from a single continuum
 Dirac operator confined to the topological sector $\nu = 0$.

\begin{figure}
 \includegraphics[width=1.0\columnwidth,clip]{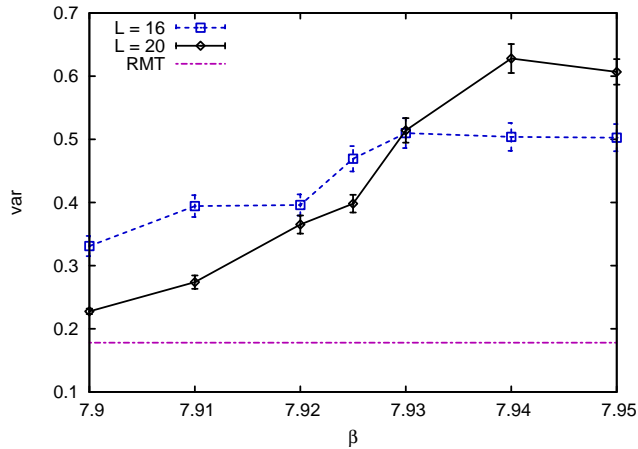}
 \caption{Level spacing variance ($\rm var$) of the low eigenvalues of
 the Dirac operator in quenched QCD for two volumes ($L^3\times4$) and values
 of the lattice coupling $\beta$ spanning the transition.
 The system undergoes a metal-insulator transition around
 $\beta_c \sim 7.93$.}  
\label{figsv} 
\end{figure}

\subsection{Eigenvalue analysis in quenched QCD}\label{secevq}

The critical temperature at which the AT close to the origin occurs 
was determined by performing a finite size scaling analysis \cite{sko}.
In essence this method consists of computing a spectral correlator
for different sizes and then seeing for what value of the temperature
it becomes scale invariant (no dependence on the size).
We recall that scale invariance is a typical signature of the AT.
Since we are interested in a small spectral 
window close to the origin we have chosen a short range
 correlator, the level spacing distribution $P(s)$ \cite{mehta}.
This is the probability of finding two neighboring eigenvalues at a distance 
$s_i = (\lambda_{i+1} - \lambda_{i})/\Delta $, with $\Delta$ being the
 local mean level spacing.
For an insulator, levels are not correlated, there is no 
 level repulsion and  $P(s) = e^{-s}$. 
In the case of a metal there is level repulsion,
 $P(s) \sim s^\alpha ~(s \ll 1)$,
 and Gaussian decay $P(s) \sim e^{-4s^2/\pi} ~ (s \gg 1)$ with $\alpha$ an
 integer depending on the 
 symmetry of the Dirac operator ($\alpha = 2$ in our case). 

In order to avoid any dependence on bin size in the spacing distribution,
 the scaling behavior of $P(s)$ is examined through its variance,
\begin{equation}
{\rm var} \equiv \langle s^2 \rangle - \langle s \rangle^2
  = \int_0^{\infty}ds ~s^2 P(s)- 1, 
\label{var}
\end{equation}
where $\langle \dots \rangle$ denotes spectral and ensemble averaging. 
The prediction for a metal (from RMT) is
  ${\rm var_{M}} \approx 0.178$ while
 an insulator (Poisson statistics) gives ${\rm var_I} = 1$.
If the variance gets closer to the metal (insulator) result as the
 volume is increased we say that the system is delocalized (localized).
Any other intermediate value of ${\rm var}$ in the thermodynamic limit
 is a signature of a metal-insulator transition.

In figure \ref{figsv} we plot the variance of the eigenvalues in the interval
 $a \lambda \in [0.025,0.05]$ (with $a$ the lattice spacing) for different
 volumes and couplings.
Increasing $\beta$ corresponds to increasing temperature.
The variance appears to be scale invariant at $\beta = 7.93$ and
 increases (decreases) with the volume for larger (smaller) $\beta$.
This behavior points to a metal-insulator transition around
 $\beta_c \sim 7.93$ for this interval.
The interval above was chosen so as to contain enough eigenvalues to obtain
  good statistics for all couplings used.
Other intervals give similar results provided that we remain sufficiently
 close to the origin although the value of $\beta_c$ may decrease
 slightly as the interval approaches zero.
Note also that as $\beta$ increases the lattice spacing $a$ decreases which,
 in turn, increases the eigenvalue range in physical units.
We have not tried to correct it in the quenched case as it would not affect
 the scaling with volume seen at fixed $\beta$ which is how the localization
 properties and $\beta_c$ are determined.
However we mention this is likely the cause of the 
 downward bend in figure \ref{figsv} for $\beta =7.95$.

\begin{figure}
 \includegraphics[width=\columnwidth,clip]{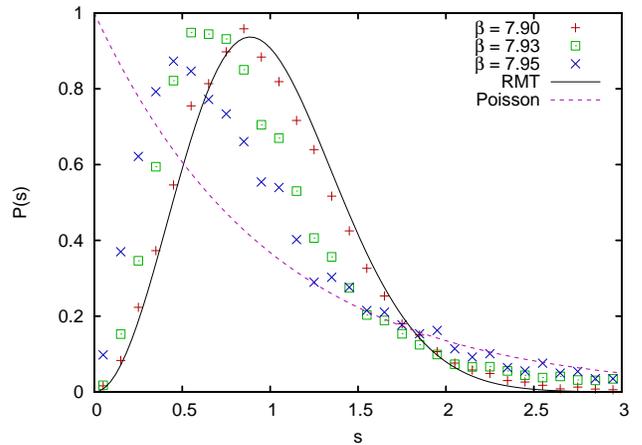}
 %\vspace{3mm}
 \caption{Level spacing distribution, $P(s)$, of the low eigenvalues of
 the Dirac operator in quenched QCD for lattice size $20^3\times4$ and values
 of the coupling around the transition.
 A transition from the random matrix prediction (RMT) typical of a metal
 towards the Poisson result typical of an insulator is observed as the
 temperature is increased (increasing $\beta$).}
\label{psnf0}
\end{figure}

Evidence of a metal-insulator transition is also seen directly
 in the level spacing distribution shown in figure \ref{psnf0}.
This is plotted for the largest volume lattices ($20^3\times4$) at different
$\beta$ along with the metal (RMT) and insulator (Poisson) results.
As the temperature is increased the distribution moves from metallic
towards an insulator similarly to the variance.
Note that, as with the variance plot, the system has still not reached the
result of an insulator at this volume, but should scale towards it as
the volume is increased.

\begin{figure}
 \includegraphics[width=1.0\columnwidth,clip]{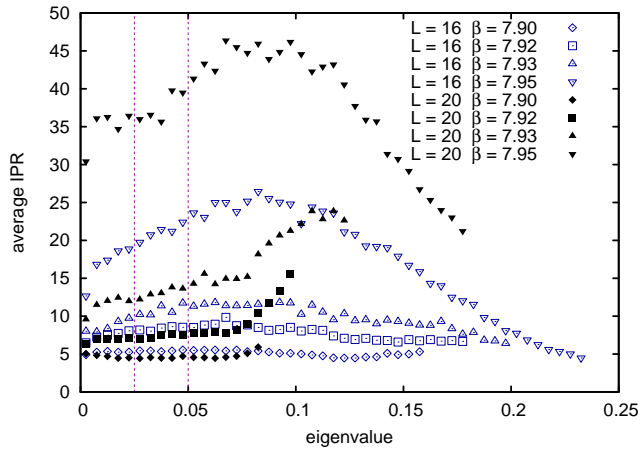}
 \caption{The average inverse participation ratio (IPR)
 versus the eigenvalue (in lattice units)
 for the low eigenvalues of
 the Dirac operator in quenched QCD for two volumes ($L^3\times4$) and values
 of the lattice coupling $\beta$ spanning the transition.
 The scaling of IPR with the volume suggests that an
 AT takes place around $\beta_c \sim 7.92 - 7.93$ (see text).
 The dotted lines show the eigenvalue range considered for locating the 
 localization transition in the quenched lattices.}
\label{fig32}
\end{figure}

\subsection{Eigenvector analysis in quenched QCD}

We now investigate whether eigenstate properties are compatible with the 
 findings of the previous section.
We recall that extended eigenstates are a signature of 
 a metal and exponential localization is typical of an insulator. 
The eigenstate decay is very sensitive to
 statistical fluctuations, boundary conditions and
 finite size effects in general
and consequently it is not the best
alternative for numerical investigations. 
A much simpler option is to study the 
scaling with volume of the eigenstate moments,
$P_q=L^{d(q-1)}\int d^{d+1} r |\psi_\lambda(r)|^{2q}$, where 
$\psi_\lambda(r)$ is a normalized eigenstate of the 
Dirac operator with eigenvalue $\lambda$ and $d$ is the 
spatial dimension. In a metallic sample it is
 expected that $P_q$ decreases with volume approaching $P_q \sim 1$
 in the thermodynamics limit.
Conversely in an insulator, $P_q$ is expected to increase 
 with volume as $P_q \propto L^{d(q-1)}$.
At the AT, eigenstates are multifractal meaning the wavefunction moments
 present anomalous scaling with respect to the sample size, 
 $P_q \propto L^{-(D_q-d)(q-1)}$, where $D_q$ is a set of exponents
 describing the transition \cite{wegner,mirlin}.
For the analysis of the lattice data we restrict ourselves
 to the second moment, $P_2$, usually referred to as the inverse participation 
 ratio (IPR).
In figure \ref{fig32} we plot the average of the IPR versus
 eigenvalue for different volumes and couplings.
We observe that for $\beta \le 7.92$ ($\ge 7.93$) the IPR decreases
 (increases) with the volume.
This suggests, in agreement with the previous spectral analysis,
 that an AT takes place around $\beta_c \sim 7.93$.
Unfortunately the range of accessible volumes is too small to provide
 a reliable estimate of the multifractal dimensions, $D_q$, in the critical
 region.

\begin{figure}
 \includegraphics[width=\columnwidth,clip]{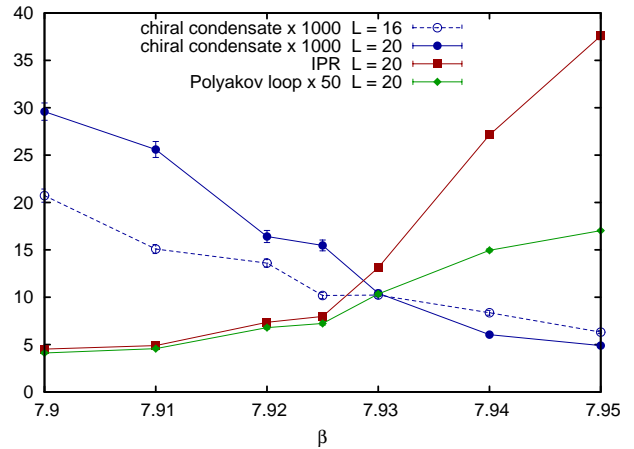}
%\vspace{3mm}
 \caption{Chiral condensate, IPR of the low eigenmodes and Polyakov loop
 in quenched QCD for lattice sizes $L^3\times4$ and values
 of the lattice coupling $\beta$ spanning the transition.
 Remarkably the localization, chiral restoration and deconfinement transitions
 all occur around the same temperature.}
\label{cciprnf0}
\end{figure}

\section{Localization and chiral restoration in quenched lattice QCD}

Having found that an AT close to the origin does occur
 in quenched QCD we now investigate its possible relation 
 with the chiral phase transition. 
Although in the quenched case there is no true chiral symmetry,
 it is still possible to study the quenched chiral condensate.
From the Banks-Casher relation (\ref{BC}) we know that the condensate
 in the chiral limit is proportional to the density of eigenvalues
 of the Dirac operator near zero.
By chiral restoration it is thus meant that the infrared limit
 of the spectral density of the Dirac operator vanishes at a
 certain temperature.
Using this definition of the condensate we can then approximate the density
 of eigenvalues near zero simply from the average of the smallest
 eigenvalue.
The precise relation can be obtained from the RMT prediction of the
 distribution of the smallest eigenvalue \cite{1ev}.
We will then compare the behavior of the quenched chiral condensate to the
 localization transition in QCD.

In figure \ref{cciprnf0} we plot the quenched chiral condensate in the
 chiral limit, the average IPR of the low modes,
 and the real part of the average Polyakov loop
 at different couplings.
Amazingly the indicators for chiral symmetry restoration,
 Anderson localization
 and deconfinement (respectively) all show signs of a transition
 around the same temperature.
We have shown the condensate for two different volumes
 so the scaling to infinite volume can be seen.
As the volume increases the condensate increases for $\beta<7.93$ and decreases
 for $\beta>7.93$ suggesting that the chiral transition occurs
 around  $\beta_c\sim7.93$.
This behavior is very similar to that found in the level spacing variance
 shown in figure \ref{figsv}.

The IPR in figure \ref{cciprnf0} is averaged over modes in the same
 eigenvalue range used in the
 previous section.  In agreement with figure \ref{fig32} we again see that
 it rises abruptly around the same $\beta_c$.
Although we can't currently provide any direct connection between the
 localization transition and deconfinement, it is still interesting to see
 that the order parameter for confinement (the Polyakov loop) shows an
 extremely similar behavior to the IPR thus suggesting that the two transitions
 are also related.
Additionally, as is shown above, we do see evidence that there is a connection between localization
and the chiral phase transition.  Indeed, the apparent agreement in critical
temperatures strongly suggests that Anderson localization is the
 mechanism driving the chiral phase transition in quenched QCD.

\begin{figure}
 \includegraphics[width=\columnwidth,clip]{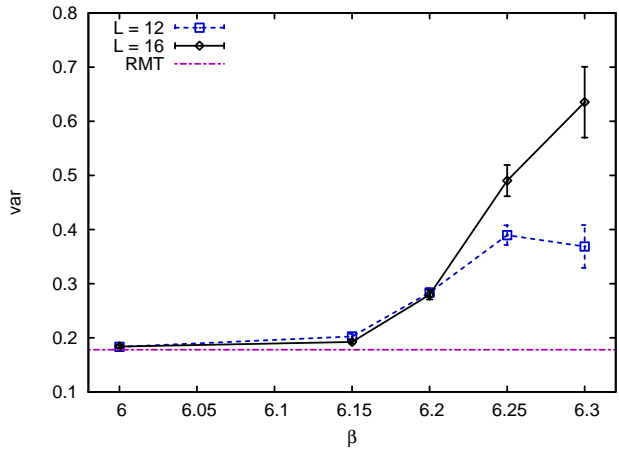}
 %\vspace{3mm}
 \caption{Level spacing variance ($\rm var$) of the low eigenvalues of
          the Dirac operator in 2+1 flavor QCD for two volumes ($L^3\times4$)
          and values of the lattice coupling $\beta$ spanning the transition.
          The system undergoes a metal-insulator transition around
          $\beta_c \sim 6.2$.}
\label{svnf21}
\end{figure}

\section{Localization and chiral restoration in full lattice QCD}
\label{secunq}

We now examine the localization transition in full lattice QCD.
With respect to the chiral transition the most important difference from
 the quenched case is
 the fact that, for nonzero quark masses, the condensate is no longer
 exclusively determined by the eigenmodes close to the origin.
The condensate now gets contributions from larger eigenvalues with a
 relative weight given in (\ref{BC}).
The importance of the modes right at zero then becomes diluted
 by the nonzero modes.
Only in the chiral limit, as the dynamical mass approaches zero, does
 the condensate only depend on the lowest modes.
Therefore in order to understand the relationship between localization and
chiral restoration we must study the localization properties over a 
 range of eigenmodes similar to the ones relevant for the condensate.
Typically this spectral window comprises not only critical eigenstates at the
 mobility edge but also localized and delocalized eigenstates around it. 
The issue of whether 
 such a mixture causes a crossover or a transition
 is subtle and requires further study.

Understanding the relationship between localization and \SB{} in full
 QCD is also complicated by the fact that at the physical quark masses
 the chiral transition is really a rapid crossover.
The critical temperature in this case is usually defined by the maximum
 of the chiral susceptibility.
However, as discussed below, the localization properties 
are related to the susceptibility and thus could still provide a connection between
 the two phenomena.

In order to explore this scenario our first task is to locate the
 mobility edge in full lattice QCD.
For this we look at the level spacing variance (\ref{var}) for two
 different lattice sizes versus coupling (figure \ref{svnf21})
in a range of eigenvalues $0\le\lambda\le30$ MeV (the relation for the 
lattice spacing in \cite{milclat} was used to convert to physical units).
This range goes up to roughly three times the light quark mass and thus
 covers an important fraction of the contribution to the condensate.
Our results are not sensitive to the length of the spectral interval utilized
 provided that there is no coexistence of localized and delocalized
 eigenstates.
If this occurs we shall observe a crossover rather than a transition.

Here we clearly see that the variance is close to the random matrix prediction 
 typical of a metal for  
 $\beta < 6.2$ while for $\beta > 6.2$ it tends rapidly to the result of
 Poisson statistics typical of an insulator.
Thus around $\beta_c \sim 6.2$ there is a transition from the
 metallic to the insulator limit.
Converting this to physical units we get $T_c \sim 195$ MeV.
By transition we really mean that there is a sharp increase of the 
variance in a quite narrow window of temperatures. 
Much larger volume would be needed to clarify whether 
it is just a crossover or a true transition.

\begin{figure}
 \includegraphics[width=\columnwidth,clip]{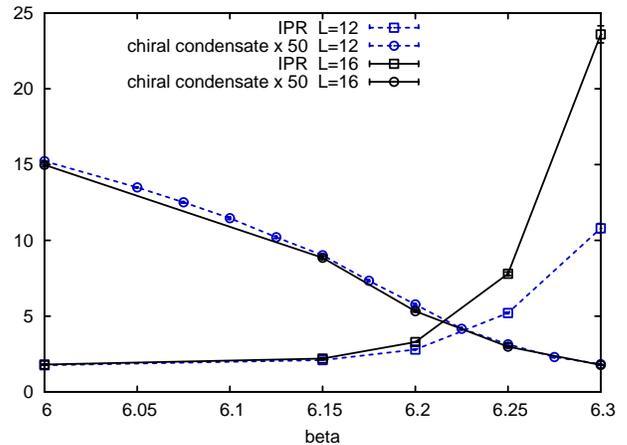}
 %\vspace{3mm}
 \caption{Chiral condensate for the light ($u,d$) quarks and
          IPR of the low eigenmodes for $2+1$ flavor QCD for two volumes
          ($L^3\times4$) and values of the lattice coupling $\beta$
          spanning the transition.
          The IPR shows signs of an AT around $\beta_c \sim 6.2$ while
          the condensate shows only a crossover.}
 \label{iprccnf21}
\end{figure} 

We see a similar scaling behavior in the IPR (figure \ref{iprccnf21}).
Again there is a transition from extended to localized states at around the
same $\beta_c \sim 6.2$.
However if we look at the chiral condensate also plotted on the same figure
we see that it does not show any volume dependence.
This is in agreement with the expected behavior at a chiral crossover.
In this case one can still define the critical temperature
from the maximum of the chiral susceptibility. For the unquenched lattices
this is about $194$ MeV \cite{ll}, which compares favorably with that
found for the AT.

The reason why a transition is observed for localization
can be understood fairly easily.
For the localization transition, a mobility edge forms at zero for some
temperature and then moves towards higher energies as the temperature
is increased.  If the mobility edge is below the spectral window being 
studied then the states will appear extended and if above then they will
appear localized.  While the mobility edge is inside the window there
will be a mix of extended, critical and localized eigenmodes.  
This will technically
give a crossover, however this is merely a consequence of the size of the
spectral window analyzed which is related to the
dynamical quark mass. From the point of view of localization 
theory the determination of what
constitutes a large and small window is governed by the rate of change
of the mobility edge with temperature which, for our case, seems to be
very fast. Thus if the window around the mobility edge is small enough a 
true transition is expected.

However the reason why we observe a crossover in the chiral transition is less clear.
The spectral region which the chiral condensate is sensitive to
 is also determined by the quark mass.
Evidently due to the nonzero mass, the chiral transition is not as
sensitive to the mobility edge itself but rather to the coexistence of 
localized, delocalized and critical eigenmodes in the spectral region of interest.
Since we find that both phenomena occur at the same temperature, it
is natural to expect that Anderson localization is still responsible for
causing the chiral transition. However 
the exact nature of the relationship is still under investigation.

In summary, our numerical results show that an AT occurs in the low energy
part of the Dirac eigenmodes at around the same temperature as the chiral
transition.
This is fully consistent with the idea of
 Anderson localization as the microscopic mechanism driving the chiral
 phase transition.
However much larger lattices and better statistics are still needed to
 fully explore this scenario, and especially to extract the
 spectrum of multifractal dimensions and critical exponents.

\section{Additional discussion on localization and chiral restoration}

This final section is devoted to a more detailed account on how
 localization and the chiral phase transition are related.
We start by addressing certain problems associated to linking these two
 phenomena.
The most serious one is the fact that the spectral density
 is not a good order parameter for the transition to localization
 since it does not vanish in any of the phases.
Signatures of the transition to localization are typically found in correlation
 functions of higher order.
By contrast the chiral condensate is directly related to the spectral density
 of the Dirac operator through the Banks-Casher relation (\ref{BC}). 
This seems to suggest
 that Anderson localization and the chiral  phase transition cannot be so
 intimately related.
However, in disordered systems with chiral symmetry (or any other
 additional discrete symmetry) the spectral density is already
 sensitive to the strength of disorder \cite{AK,ant1} so it may still
 play the role of an order parameter for the transition.
Additionally the fluctuations of the order parameter are related to
 density-density correlations which are sensitive to localization
 effects even in systems with no chiral symmetry.

Another issue that deserves further clarification is the choice of the
term Anderson localization. In the context of disordered systems, this
term is used if localization is produced by destructive
interference. On the other hand the term Mott transition refers to a
transition caused by interactions. Quark interactions are a key
ingredient in QCD so it may seem more appropriate to use Mott-Anderson instead
of Anderson. We stick to Anderson to emphasize that disorder, due to
the fluctuations of gauge fields, plays a crucial role in the
transition as well.  An indication that this is the case is the fact
that even for temperatures just above chiral restoration, QCD
is still non-perturbative \cite{SZ} thus suggesting the mechanism driving the
transition is not exclusively a weakening of the strong interactions.

Finally we discuss whether the order and critical exponents of the
 phase transition can be deduced from the study of localization.
 In the case of the standard non-chiral AT evidence that some
 sort of phase transition takes places is the fact that the
 localization length diverges at the transition with a certain
 universal critical exponent $\nu$. Also the spectrum of
 multifractal dimensions $D_q$ at the AT must be related to certain
 critical exponents of the chiral phase transition.  For instance, the
 susceptibility is given by the integral of a density-density
 correlation function whose decay is controlled by the multifractal
 dimension $D_2$ \cite{kravtsov}.

We recall that critical exponents related to $D_q$ cannot be predicted
by any mean field theory even if perturbative corrections are taken
into account.  Unfortunately the range of volumes used here
is still too small to provide a reliable estimate of the
multifractal dimensions $D_q$ in the critical region and its relation
to critical exponents.  However in the context of ILM \cite{aj1} it
has already shown that, at the chiral phase transition, eigenvectors
of the QCD Dirac operator are multifractal with a $D_q$ similar to the
one found in a 3D disordered conductor at the AT.  This is another
indication that Anderson localization is an important ingredient to understand the
chiral phase transition in QCD.

In conclusion, we have studied Anderson localization of the QCD Dirac
operator at nonzero temperature.  Near the origin we found a clear
transition from delocalized to localized states as the temperature is
increased.  Around this mobility edge, the eigenvectors and spectral
correlations are similar to those of a disordered system undergoing an
AT.  Remarkably both the transition to localization and the chiral
phase transition 
occur at the same temperature.  This indicates that the phenomenon of
Anderson localization plays a crucial role in the restoration of the chiral 
symmetry.

AMG thanks C. Gattringer, M. Teper and B. Bringoltz for illuminating
discussions.
JCO thanks C. DeTar and L. Levkova for data and discussions pertaining to the
MILC lattices.
AMG was supported by a Marie Curie Outgoing Fellowship,
contract MOIF-CT-2005-007300.
JCO was supported in part by U.S. DOE grant DE-FC02-01ER41180.
The computations were performed at FNAL and BU.


\vspace{-6mm}
\begin{thebibliography}{9}
\vspace{-6mm}

\bibitem{callan} C.G. Callan, R. Dashen and D.J. Gross,
 Phys. Rev. D {\bf 17}, 2717 (1978).

\bibitem{diakonov} D. Diakonov and V. Petrov,
 Nucl. Phys. {\bf B245}, 259 (1984).

\bibitem{shuryak} E. Shuryak, Nucl. Phys. {\bf B203}, 93,116,140 (1982).

\bibitem{polyakov} A. Belavin, A. Polyakov, A. Schwartz and Y. Tyupkin,
 Phys. Lett. {\bf 59}, 85 (1975).

\bibitem{thooft} G. 't Hooft, Phys. Rev. Lett. {\bf 37}, 8 (1976).

\bibitem{bank} T. Banks and A. Casher, Nucl. Phys. {\bf B169}, 103 (1980).

\bibitem{SS97} T. Sch\"afer and E. Shuryak,
 Rev. Mod. Phys. {\bf 70}, 323 (1998).

\bibitem{wilczek} R.D. Pisarski and F. Wilczek,
 Phys. Rev. D {\bf 29}, 338 (1984).

\bibitem{revchi} H. Meyer-Ortmanns, Rev. Mod. Phys. {\bf 68}, 473 (1996).

\bibitem{mendes} T. Mendes, hep-lat/0609035.

\bibitem{anderson} P.W. Anderson, Phys. Rev. {\bf 109}, 1492 (1958).

\bibitem{guhr} T. Guhr, A. Mueller-Groeling and H.A. Weidenmueller,
 Phys. Rept. {\bf 299}, 189 (1998).

\bibitem{locgw}
 R.G. Edwards, U.M. Heller, J. Kiskis and R. Narayanan,
  Phys. Rev. Lett. {\bf 82}, 4188 (1999);
 C. Gattringer, M. G\"ockeler, P.E.L. Rakow, S. Schaefer and A. Sch\"afer,
  Nucl. Phys. {\bf B618}, 205 (2001);
 R.V. Gavai, S. Gupta and R. Lacaze, Phys. Rev. D {\bf 65}, 094504 (2002);
 M. Golterman and Y. Shamir, Phys. Rev. D {\bf 68}, 074501 (2003);
 M. Golterman, Y. Shamir and B. Svetitsky,
  Phys. Rev. D {\bf 72}, 034501 (2005);
 M.I. Polikarpov, F.V. Gubarev, S.M. Morozov and V.I. Zakharov,
  PoS {\bf LAT2005}, 143 (2005);
 V. Weinberg, E.-M. Ilgenfritz, K. Koller, Y. Koma, G. Schierholz and
  T. Streuer, PoS {\bf LAT2005}, 171 (2005);
 Y. Koma, E.-M. Ilgenfritz, K. Koller, G. Schierholz, T. Streuer and
  V. Weinberg, PoS {\bf LAT2005}, 300 (2005);
 E.-M. Ilgenfritz, K. Koller, Y. Koma, G. Schierholz, T. Streuer and
  V. Weinberg, Nucl. Phys. Proc. Suppl. {\bf 153}, 328 (2006).

\bibitem{locstag}
 P.H. Damgaard, U.M. Heller and A. Krasnitz,
  Phys. Lett. B {\bf 445}, 366 (1999);
 F. Farchioni, Ph. de Forcrand, I. Hip, C. B. Lang and K. Splittorff,
  Phys. Rev. D {\bf 62}, 014503 (2000);
 P.H. Damgaard, U.M. Heller, R. Niclasen and K. Rummukainen,
  Nucl. Phys. {\bf B583}, 347 (2000);
 M. G\"ockeler, P.E.L. Rakow, A. Sch\"afer, W. S\"oldner and T. Wettig,
  Phys. Rev. Lett. {\bf 87}, 042001 (2001);
 C. Bernard, \emph{et. al.}, PoS {\bf LAT2005}, 299 (2005).

\bibitem{locother}
 J. Greensite, S. Olejnik, M.I. Polikarpov, S.N. Syritsyn and V.I. Zakharov,
  PoS {\bf LAT2005}, 325 (2005);
 J. Greensite, A.V. Kovalenko, S. Olejnik, M.I. Polikarpov, S.N. Syritsyn
  and V.I. Zakharov, Phys. Rev. D {\bf 74}, 094507 (2006);
 Z. Fodor and S.D. Katz, JHEP {\bf 0404}, 050 (2004).

\bibitem{diakonov1} D. Diakonov and P. Petrov,
 Phys. Lett. B {\bf 147}, 351 (1984);
 Sov. Phys. JETP {\bf 62}, 431 (1985);
 Nucl. Phys. {\bf B272}, 457 (1986).

\bibitem{VO} J.C. Osborn and J.J.M. Verbaarschot,
 Phys. Rev. Lett. {\bf 81}, 268 (1998);
 Nucl. Phys. {\bf B525}, 738 (1998).

\bibitem{zahed} R.A. Janik, M.A. Nowak, G. Papp and I. Zahed,
 Phys. Rev. Lett. {\bf 81}, 264 (1998).

\bibitem{aj} A.M. Garcia-Garcia and J.C. Osborn,
 Phys. Rev. Lett. {\bf 93}, 132002 (2004).

\bibitem{aj1} A.M. Garcia-Garcia and J.C. Osborn,
 Nucl. Phys. {\bf A770}, 141 (2006).

\bibitem{diakolet} D. Diakonov, hep-ph/9602375.

\bibitem{sym} K. Symanzik, {\it Recent developments in gauge theories},
 eds. G. `t Hooft, \emph{et. al.}, Plenum, New York, 313 (1980).

\bibitem{asqtad}  K. Orginos and D. Toussaint, 
  Phys. Rev. {\bf D59}, 014501 (1999); 
  Nucl. Phys. (Proc. Suppl.) {\bf 73}, 909 (1999);
  G. P. Lepage, Nucl. Phys. (Proc. Suppl.) {\bf 60A}, 267 (1998);
  Phys. Rev. D {\bf 59}, 074502 (1999).

\bibitem{milclat} C. Bernard, \emph{et. al.}, PoS {\bf LAT2005}, 156 (2005).

\bibitem{sko} B.I. Shklovskii, \emph{et. al.},
  Phys. Rev. B {\bf 47}, 11487 (1993).

\bibitem{mehta} M.L. Mehta, {\it Random Matrices},
  Academic Press, New York, 2nd edition (1991).  

\bibitem{wegner} F. Wegner, Z. Phys. B {\bf 36}, 209 (1980);
  B.L. Altshuler, V.E. Kravtsov and I.V. Lerner,
  \emph{Mesoscopic Phenomena in Solids},
  eds. B.L. Altshuler, \emph{et. al.}, North Holland, Amsterdam (1991);
  V.I. Falko and K.B. Efetov, Europhys. Lett. {\bf 32}, 627 (1995).

\bibitem{mirlin} A.D. Mirlin, \emph{et. al.},
  Phys. Rev. E {\bf 54}, 3221 (1996);
  F. Evers and A.D. Mirlin, Phys. Rev. Lett. {\bf 84}, 3690 (2000);
  E. Cuevas, \emph{et.al.}, Phys. Rev. Lett. {\bf 88}, 016401 (2002).

\bibitem{1ev} S. M. Nishigaki, P. H. Damgaard and T. Wettig,
  Phys. Rev. D {\bf 58}, 087704 (1998).

\bibitem{ll} L. Levkova (MILC Collaboration), private communication.

\bibitem{AK}  A.M. Garc\'{\i}a-Garc\'{\i}a and K. Takahashi,
  Nucl. Phys. {\bf B700}, 361 (2004); 
  A. Parshin and H.R. Schober, Phys. Rev. B {\bf 57}, 10232 (1998).

\bibitem{ant1} A.M. Garcia-Garcia and J.J.M. Verbaarschot,
  Nucl. Phys. {\bf B586}, 668 (2000).

\bibitem{SZ} E.V. Shuryak and I. Zahed, Phys. Rev. C {\bf 70}, 021901 (2004).

\bibitem{kravtsov} V.E. Kravtsov and K.A. Muttalib,
  Phys. Rev. Lett. {\bf 79}, 1913 (1997);
  S. Nishigaki, Phys. Rev. E {\bf 59}, 2853 (1999).

\end{thebibliography}
\end{document}